# Feasibility of using LSPR on the biased nanotip to realize the atomic-resolution near-field optical detection

Gong Li Hui[a]

*Department of Physics, Guangxi Normal University, Guilin, 541004, China.*

A near-field optical detection method and its theoretical model are developed, which follow from an algebra-based conclusion: in the localized surface plasmon resonance (LSPR) region, the extinction coefficient of a metal nanosphere is the sesquiplicate proportion of the localized electron density. Eleven spectral tests of gold, silver, and aluminum nanospheres are used to verify this model. For a metal nanosphere, the frequency and intensity of the LSPR are dependent on the localized electron density only. The electron density can be tuned by adjusting the bias voltage, so the bias can enhance or inhibit the LSPR frequency of the metal nanosphere.

**I. Introduction**

The optical microscope forms a part of the basis of the nanotechnology. Today, our continuously increasing ability to construct nanomaterials has led to an urgent need for new optical microscope with atomic-resolution .

It is a well-known fact that the applied bias voltage affects the resolution of scanning tunneling microscopes (STM) [1-4]. To study this phenomenon in the case of a scanning near-field optical microscope (SNOM), a modified Lorentz model has been employed [5, 6]. This approach has illustrated that adjustments to the bias can improve the near-field detection resolution, leading to the development of a theoretical model and a patented [7] detection method.

Nano-metal enhancement of the near-field has been examined extensively, particularly over the past decade [8, 9], and a series of promising discoveries have been made [10-16]. This paper presents the patented [7] near-field detection methods shown in Fig. **1** and clarifies the mechanism through which adjustments to the bias voltage improve the near-field detection resolution.

---

[a] lllren@sina.com; Current affiliation: Department of Physics, Guangxi Normal University, Guilin, 541004, China.

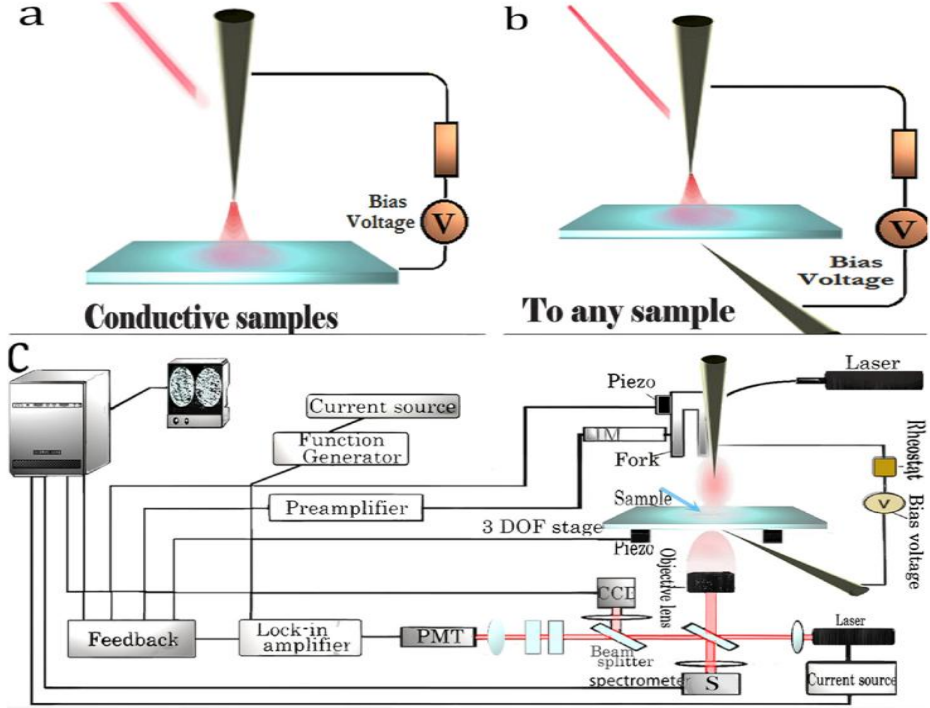

**FIG. 1.** (a), (b) Proposed near-field detection method; (c) Scanning near-field optical microscope (SNOM) system based on proposed method.

## II. THEORETICAL MODEL AND EXPERIMENTS

If the TM-polarized incident laser has frequency ω, the electric dipole moment of a metal tip under laser illumination in air is defined as

$$\tilde{P} = nq(x_0 e^{i\omega t}) = (\epsilon_\omega - 1) E e^{i\omega t}, \quad (1)$$

where $x_0$ is the amplitude of electron resonance, $\epsilon_\omega$ is the dielectric constant, which expressed as $\epsilon_\omega = \epsilon_{\omega 1} + i\epsilon_{\omega 2}$. q is single-electron charge and n is the number of localized electrons, and E is the electric field of the incident waves. According to the dipole antenna model, the distribution of the near-zone electric field can be expressed in spherical coordinates as

$$\vec{E} \cong \frac{\tilde{P}}{4\pi R^3 \varepsilon_0} \left(2\hat{E}_R \cos\theta + \hat{E}_\theta \sin\theta\right). \quad (2)$$

The tip is considered simply as a nanosphere of radius r, where $r \ll \lambda$ (the laser wavelength). Then, the relationship between the electron distribution and electric field V can be expressed as

$$N(\theta) = 3V \cos\theta. \quad (3)$$

A single oscillator model incorporating radiative damping has been shown to accurately capture the spectral features of metal [5, 6, 18]. According to this model, the electron motion of a noble metal can be expressed as [17]

$$\tilde{x} = -q\tilde{E}/[m_e(-\omega^2 + i\gamma\omega + \omega_0{}^2)], \quad (4)$$

where the damping is given by $\gamma\dot{x} = \gamma_a\dot{x} + \gamma_s\dddot{x}$, in addition to the internal damping force $\gamma_a\dot{x}$, the charge experiences an additional force $\gamma_s\dddot{x}$ due to radiation reaction, where $\gamma_s = q^2/6\pi c^3$ [5, 6, 18]. Localized surface plasmon resonance (LSPR) is generally considered to occur at the Froehlich frequency in the region where $\epsilon_{\omega 1} = -2$ and $\epsilon_{\omega 1} \approx 0$, although this is not strictly correct [19].

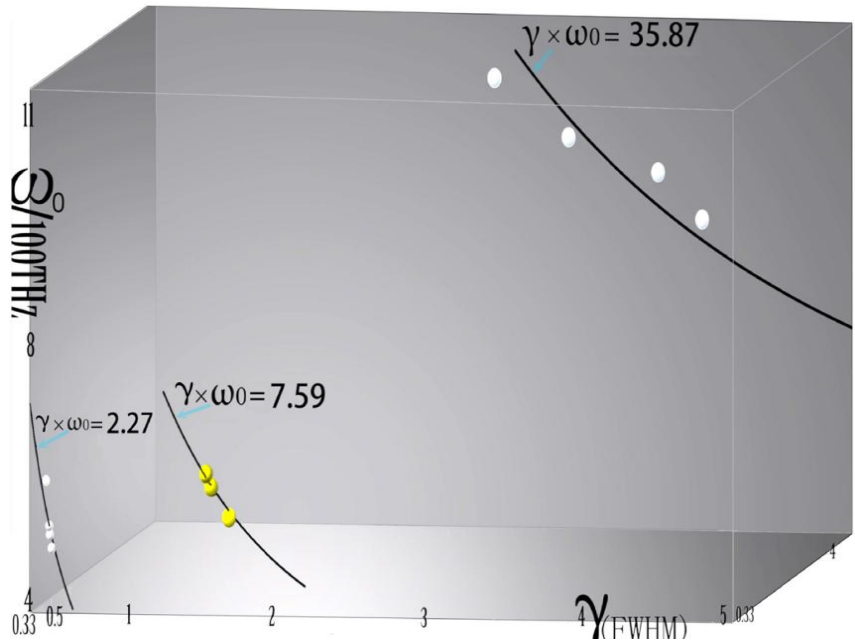

**FIG. 2.** Data from equation (5) is compared with ultraviolet-visible spectroscopy (UV-Vis) spectral data. Theoretical hyperbolae and experimental data are presented in order from left to right, and correspond to silver, gold, and aluminum nanospheres.

Substituting $\epsilon_{\omega 1} = -2$ into equation (1), which contains two independent equations, yields two new equations. If these equations are then combined with equation (4), we can obtain a concise equation: $\omega_p^2 \cong 3\gamma\omega_0$. Note that some constants are omitted, so a more accurate form of this expression is

$$\gamma \propto \omega_0{}^{-1}. \quad (5)$$

This work may be the first to take note of and examine equation (5). The accuracy of equation (5) has been examined and confirmed for a variety of cases, i.e., a total of eleven spectral tests of differently sized silver, aluminum, and gold nanospheres. The results are shown in Fig. 2.

It must be emphasized that $\gamma$ can only be equal to the half-peak width in a normalized extinction curve. The dielectric constant is expressed as $\epsilon_\omega = \epsilon_{\omega 1} + i\epsilon_{\omega 2}$, according to Mie theory, and the extinction coefficient $K$[19, 20] is

$$K = \frac{18\pi f \epsilon_0^{\frac{3}{2}}}{\lambda} \left[ \frac{\epsilon_{\omega 2}}{(\epsilon_{\omega 1} + 2\epsilon_0)^2 + \epsilon_{\omega 2}^2} \right], \quad (6)$$

where the volume factor $f = N \times V$, with $V$ being volume and $N$ atomic density. For metal, the number of valence electrons multiplied by $N$ is the electron density $N_e$. In the region of the LSPR central frequency, equation (5) can be substituted into equation (6) to yield

$$K = \frac{24\pi^2 r^3 N}{C} \left[ \frac{\gamma \omega_0^2}{\omega_p^2} \right] \propto N_e \omega_0. \quad (7)$$

Here, $\omega_0$ and $N_e^{1/2}$ are approximately proportional; therefore, the relationship between the near-field intensity and electron density of the metal tip can be approximated as

$$I \propto K \propto N_e^{3/2}. \quad (8)$$

Note that this report is the first to obtain and study equation (8). Equations (3) and (8) demonstrate that the bias can both enhance and concentrate the near field. The data calculated from equation (8) is highly consistent with data reported in Ref. [21] for sodium spectra, as shown in Fig. 3.

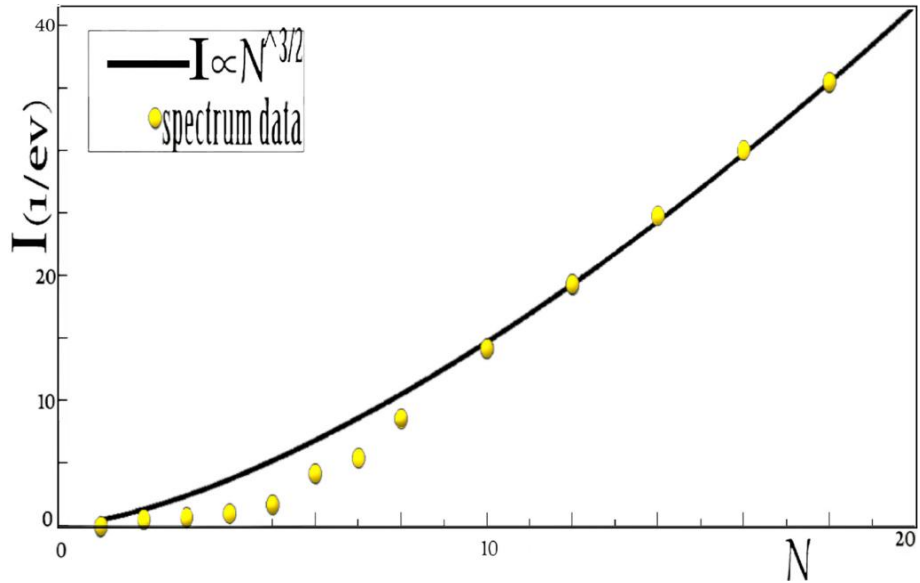

**FIG. 3.** Equation (8) is highly consistent with LSPR spectral data, for a varying number of atoms.

By combining equations (2), (3), (7), and (8), this report obtains an accurate model that can describe the near-field strength distribution before and after a bias voltage is applied to the probe, with

$$I(R,\theta,\varphi) \propto \frac{(N_0+3V\cos\theta)^{3/2}}{\varepsilon_m (R/r)^3}, \qquad (9)$$

where $N_0$ is the electron density before application of the bias voltage and $\varepsilon_m$ is the dielectric constant of the environment. Further, the electron density is uneven after application of the bias, so the LSPR frequency of each zenith angle is changed and is expressed as $\omega_0(\theta)$. For an incident laser with the frequency of the tip-end LSPR, the phase shift $\Delta\Phi$ is related to $\omega_0(\theta)^2 - \omega_0(0)^2$, but surprisingly, this is not contained in equation (9). However, in a near-field detection scenario such as that shown in Fig. 4, the tip radius is ten times the distance, and the amplitude variation with angle can be neglected. Then, we obtain

$$\frac{I(\theta)}{I_{max}} = \left(\frac{\sin\Delta\Phi/2}{\Delta\Phi/2}\right)^2 \cong \left(\frac{0.207}{0.209}\right)^2. \qquad (10)$$

So, the phase shift due to bias has little effect on the near-field equipment. Therefore, it can be said that equation (9) is rigorous.

## III. SIMULATION

Based on equation (9), we simulate the near field of an SNOM before and after a bias voltage is applied to the probe, using MATLAB 2014a (in the region of the LSPR frequency). The results are shown in Fig. 4.

After a bias voltage is applied to the probe, the simulation in Fig. 4 shows that the near-field is not only enhanced significantly, but is also concentrated toward the front-most end of the tip. According to the uncertainty principle, field enhancement corresponds to higher vertical resolution of the near-field detection, while field concentration suggests higher lateral resolution.

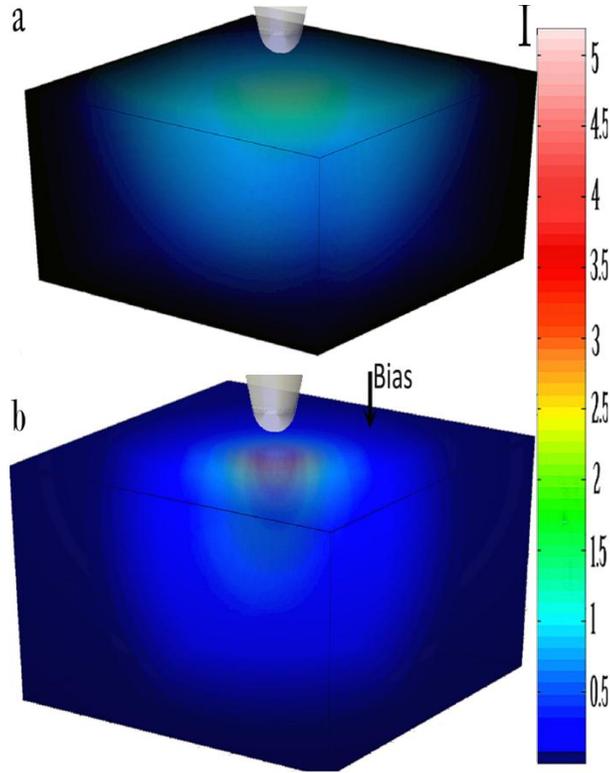

**FIG. 4.** Near-field distribution in 3D-space. The tip radius is 50 nm, the section is 5 nm from the tip, and the incident wave is TM polarized. (a) Before and (b) after probe is biased. The near-field is stronger and accumulated in the axis direction in (b).

## IV. DISCUSSION

In this paper, the expression of the extinction coefficient of the metal nano-tip was simplified significantly by Eq. (5). As shown in Fig. 2, Eq. (5) exhibits high accuracy in comparison with experimental data.

Eq. (8) is an inference of Eq. (5) and, in Fig. 3, it is compared with previously reported data as a supplementary evaluation, to increase the objectivity of this study.

This report reveals a phenomenon: in the region of the LSPR frequency of a metal nanosphere, the exciting field is the sesquiplicate proportion of the localized electron density. This explains the benefits of designs such as that shown in Fig. 1, which quantify the near-field optical signal used for sample characterization more easily. In detail, the benefit of proposed system is based on the following principles:

1) Bias voltage is applied to the tip in order to tune the electron density, to improve the optical contrast and clarity of near-field detections.

2) The most simplified metal-tip extinction coefficient must ideally be controlled, reproducible for a standard procedure, and have broad applicability.

3) By fortunate coincidence, the simplest mathematical model of the near-field at LSPR frequency yields the highest frequency S/N (signal to noise ratio). As a bias voltage can be used to tune the localized electron states on the tip surface as required, this approach could aid in the realization of high-volume manufacturing of tips with high field enhancement.

All these aspects of near-field detection continue to pose a challenge. Therefore, this work illustrates that a design such as that shown in Fig. **1** allows atomic-resolution near-field optical technology to be realized. The main limitation of this research is that the proposed design has not been verified for a SNOM. Unrecognized errors may exist in this work due to the author's lack of relevant experience.

## V. CONCLUSION

This work can be extended to rotationally symmetrical metal nanoparticles in the axial direction. Equations (**5**), (**8**), and (**9**) are the original findings of this paper. In particular, equation (**9**) models the near-field distribution before and after a bias voltage is applied to a metal. These equations are shown to be highly consistent with experimental data obtained in this work and in Ref. [**21**], and indicate the broad and accurate applicability of this approach to describing the LSPR of metals, including noble metals. This report indicates that bias-voltage application can be used to realize atomic-resolution near-field optical technology.

**ACKNOWLEDGMENTS**

The author is grateful to Prof. Wang Li-Hu for fruitful discussions and providing experimental insights. We thank the University of Nantong for experimental resources.